\documentclass[aps,prl,twocolumn,showpacs,groupedaddress]{revtex4}
\usepackage{graphicx}
\usepackage{dcolumn}
\usepackage{bm}
\usepackage{amssymb}

\newcommand{\met}{\mbox{$\;\!\not\!\! E_{T}$}}
\newcommand{\gev}{\,\mathrm{GeV}}
\newcommand{\tev}{\,\mathrm{TeV}}
\newcommand{\fbinv}{\,\mathrm{fb}^{-1}}

\begin{document}

\hspace{5.2in} \mbox{FERMILAB-PUB-08/334-E}

\title{Search for pair production of second generation scalar leptoquarks 
}
%
\author{V.M.~Abazov$^{36}$}
\author{B.~Abbott$^{75}$}
\author{M.~Abolins$^{65}$}
\author{B.S.~Acharya$^{29}$}
\author{M.~Adams$^{51}$}
\author{T.~Adams$^{49}$}
\author{E.~Aguilo$^{6}$}
\author{M.~Ahsan$^{59}$}
\author{G.D.~Alexeev$^{36}$}
\author{G.~Alkhazov$^{40}$}
% ...

\affiliation{\vspace{0.1 in}(The D\O\ Collaboration)\vspace{0.1 in}}
% ...
\affiliation{$^{6}$University of Alberta, Edmonton, Alberta, Canada,
                Simon Fraser University, Burnaby, British Columbia, Canada,
                York University, Toronto, Ontario, Canada, and
                McGill University, Montreal, Quebec, Canada}
% ...
\affiliation{$^{29}$Tata Institute of Fundamental Research, Mumbai, India}
% ...
\affiliation{$^{36}$Joint Institute for Nuclear Research, Dubna, Russia}
% ...
\affiliation{$^{40}$Petersburg Nuclear Physics Institute,
                St. Petersburg, Russia}
% ...
\affiliation{$^{49}$Florida State University, Tallahassee, Florida 32306, USA}
% ...
\affiliation{$^{51}$University of Illinois at Chicago,
                Chicago, Illinois 60607, USA}
% ...
\affiliation{$^{59}$Kansas State University, Manhattan, Kansas 66506, USA}
% ...
\affiliation{$^{65}$Michigan State University,
                East Lansing, Michigan 48824, USA}
% ...
\affiliation{$^{75}$University of Oklahoma, Norman, Oklahoma 73019, USA}
% ...
 
\date{August 29, 2008}

\begin{abstract}
We report on a search for the pair production of second generation
scalar leptoquarks ($LQ$) in $p\overline{p}$ collisions at 
the center of mass energy $\sqrt{s}=1.96\,$TeV using a data set 
corresponding to an integrated luminosity of $1.0\,\mathrm{fb}^{-1}$ collected
with the D0 experiment at the Fermilab Tevatron Collider.
Topologies arising from the
$LQ\overline{LQ}\rightarrow\mu q\nu q$ and 
$LQ\overline{LQ}\rightarrow\mu q\mu q$ decay modes are investigated.
No excess of data over the standard model prediction is observed and
upper limits on the leptoquark pair production cross section are
derived at the $95\%$~C.L. as a function of the leptoquark mass 
and the branching fraction $\beta$ for the decay $LQ\rightarrow\mu q$. 
These are 
interpreted as lower limits on the leptoquark mass as a function of $\beta$.
For $\beta=1\,(0.5)$, scalar
second generation leptoquarks with masses up to 316\,GeV (270\,GeV) are
excluded.
\end{abstract}

\maketitle 

%
%
The observed symmetry between lepton and quark generations could 
be explained by new gauge bosons introducing couplings between the
lepton and quark sectors. Such particles, commonly referred to as 
leptoquarks~\cite{lqth}, would carry both lepton and baryon quantum numbers
as well as fractional electric charge.
Extensions of the standard model (SM) based on a larger gauge symmetry group 
usually predict the existence of massive leptoquarks.
Experimental bounds on lepton number violation, proton decay, 
and flavor changing neutral currents constrain hypothetical leptoquarks with 
masses of several hundred GeV to a few TeV to 
couple only to one quark and one lepton family, via processes 
conserving both lepton and baryon quantum numbers. 
Three generations
of leptoquarks can thus be 
distinguished by considering the lepton observed in the leptoquark decay.

At the Fermilab Tevatron $p\bar{p}$ Collider, leptoquarks would 
predominantly be 
produced in pairs via $q\bar{q}$ annihilation into a gluon in the $s$-channel 
independently of the unknown coupling $\lambda$ between the leptoquark and its 
associated lepton and quark. Thus, for scalar leptoquarks
the production cross section depends only on the strong coupling constant 
and the assumed leptoquark mass. 
The additional contribution to leptoquark pair production 
from $t$-channel lepton exchange with a cross 
section proportional to $\lambda^2$ can be neglected. 
When adopting the assumption that the leptoquarks couple to 
leptons and quarks of the same generation, the $t$-channel process is further
suppressed
for second and third generation leptoquarks
due to the vanishing parton distribution functions (PDFs) at high 
proton momentum fractions $x$ for partons other than $u$ and $d$ quarks.
Leptoquarks can decay either into a charged lepton and a quark with a branching
fraction $\beta$ or
into a neutrino and quark with a branching fraction $(1-\beta)$, assuming that
the leptoquark mass is much larger than the masses of its decay products, which
is generally the case for first and second generation leptoquarks.
Consequently leptoquark pair production could lead to
three characteristic final states: $\ell^+ q \ell^- q$, $\ell^\pm q \nu q$, 
and $\nu q\nu q$, with branching fractions $\beta^2$, $2\beta(1-\beta)$, and
$(1-\beta)^2$, respectively.

This Letter describes a search for second
generation scalar leptoquark pair production in the decay modes
$LQ\overline{LQ}\rightarrow\mu q \nu q$ 
and $LQ\overline{LQ}\rightarrow\mu q \mu q$ using $p\bar{p}$ collisions at  
the center of mass energy $\sqrt{s}=1.96\,$TeV recorded with the D0 detector.
These channels lead to 
topologies with
one muon, missing transverse energy (from which the transverse 
momentum of the neutrino is inferred), and two jets ($\mu\met jj$ signature), 
or with two muons and two jets  ($\mu\mu jj$ signature), respectively. 
At Run\,II of the Tevatron, the dimuon signal was previously studied by the 
D0 collaboration~\cite{d0lq2}
while the CDF collaboration studied both dimuon and single muon
signals~\cite{cdflq2} with
smaller data sets.
Since one of the muons of the $LQ\overline{LQ}\rightarrow\mu q \mu q$
decay mode might not be reconstructed, this signal contributes to the 
single muon signature as well, which is taken into account in our analysis.
The contribution of $LQ\overline{LQ}\rightarrow\mu q \nu q$ in the 
dimuon selection can be neglected due to the small probability
for a jet to mimic an isolated muon.

The signal sensitivity for both signatures depends on $\beta$.
The branching fractions for the decay modes 
$LQ\overline{LQ}\rightarrow\mu q\mu q$ and 
$LQ\overline{LQ}\rightarrow\mu q\nu q$ are 
maximal at $\beta=1$ and $\beta=0.5$, respectively. The branching fraction
of both decay modes vanishes for $\beta=0$, where the leptoquark
would decay exclusively into a neutrino and quark. The resulting
acoplanar jet topology has not been investigated in this analysis, but
was studied recently using a larger data set~\cite{d0lqaco2}.

%
%
The D0 detector~\cite{run2det} is designed to maximize the detection
and identification of 
particles arising from $p\bar{p}$ interactions and is constructed of 
dedicated subsystems arranged around the interaction point.
The central tracking system, located at the innermost part of the detector, 
consists of a silicon microstrip 
tracker (SMT) and a central fiber tracker (CFT) which cover the 
pseudorapidity regions $|\eta|<3$ and $|\eta|<2.5$, respectively. 
The pseudorapidity is defined
as $\eta = -\ln[\tan(\theta/2)]$ where $\theta$ is the polar angle with the
proton beam.
A  2\,T 
superconducting solenoidal magnet is positioned between the 
tracking system and the surrounding central and forward preshower detectors.
The calorimeter is composed of a central section (CC) which covers 
$|\eta|\lesssim 1.1$ and two end calorimeters (EC) 
that extend coverage to $|\eta|\approx 4.2$. The three calorimeter sections
are housed in their own cryostats and consist of successive layers of 
mostly uranium absorbers and active liquid argon~\cite{run1det}. 
The muon system~\cite{run2muon} is located outside the calorimeter
and covers the
region $|\eta|<2$. It consists of a layer of drift tubes and scintillation 
counters before 1.8\,T iron toroids and two similar layers outside the 
magnets.

%
%
This search for leptoquark pair production is based on an integrated 
luminosity of $1.0\,$fb$^{-1}$. 
The data samples for the single muon and dimuon analyses are 
selected with combinations of single muon triggers~\cite{pcthesis}.

%
%
Muons are identified in the region $|\eta|<2$ using track segments 
found in the muon detector which are required to have hits in both the drift
tubes and the scintillation counters. The segments are matched to tracks
reconstructed in the central tracking system which determine the muon momenta.
Muons with a transverse momentum $p_T>20\,\mathrm{GeV}$ are kept.  
A veto on cosmic ray muons is applied which is based on 
timing information in the muon system and the removal of events with an
apparent muon pair back-to-back in pseudorapidity.
For the $\mu\met jj$ selection, a tight muon identification is 
applied in order to suppress events without prompt muons. Tight quality
is defined by requiring additional hits 
in the muon detector and
that the muon track be isolated from
other tracks  
with the sum of the transverse momenta of all other tracks in a cone defined 
in terms of
$\eta$ and azimuth $\phi$ with
radius $\mathcal{R} = \sqrt{(\Delta\eta)^2+(\Delta\phi)^2}<0.5$ around the muon
less than 2.5\,GeV~\cite{pcthesis}. The muon isolation is further improved
by selecting events in which 
the energy measured in the calorimeter in a hollow cone of radius 
$0.1<\mathcal{R}<0.4$ around the muon is less than 2.5\,GeV.
In the case of the $\mu\mu jj$ selection, where the existence of two 
reconstructed muons allows for a better separation of signal and background, 
the hit requirement is loosened and only a track isolation criterion with a 
threshold of 4\,GeV on the transverse momentum sum is required
(loose quality).

Jets are reconstructed with an iterative, midpoint cone algorithm 
with a cone radius of $0.5$~\cite{d0jets}. Only jets found 
within $|\eta|<2.5$ and with $p_T>25\,\mathrm{GeV}$ are 
kept.
The jet energies are
calibrated as a function of the jet transverse energy and
$\eta$~\cite{d0incljet}. 

The transverse momentum of a final state neutrino can be inferred from the 
$\met$, calculated as the vector sum of the transverse energies in
the calorimeter cells which is corrected with the
transverse momenta of the selected muons and the jet energy calibration.

%
%
The main SM background to the
pair production of leptoquarks followed by their decay into the
$\mu q\nu q$ and $\mu q\mu q$ final states is the associated production 
of jets with $W$ and $Z/\gamma^{*}$ bosons, respectively.
Vector boson production is simulated using the \textsc{alpgen}
event generator~\cite{alpgen} which is interfaced to 
\textsc{pythia}~\cite{pythia} for the simulation of parton showering and 
hadronization.
Samples with up to five or three 
partons in addition to the $W$ or $Z/\gamma^*$ boson, respectively, are
generated and combined using the MLM matching prescription~\cite{mlm}.
For these samples as well as for the ones described below, the CTEQ6L1 PDF
sets~\cite{cteq} are used.
Additional samples for $W$ and $Z/\gamma^{*}$ boson production are generated 
with the \textsc{pythia} event generator and 
utilized to determine uncertainties in the transverse momentum shape
of the associated jets.
Top quark pair production also contributes to 
the analyzed final states and is simulated with \textsc{pythia} 
assuming a top quark mass of 175\,GeV. The multijet background, in which
a jet is misidentified as an isolated muon, is estimated from data for
the $\mu\met jj$ selection by inverting the muon 
isolation requirement and normalizing the obtained rate to data with standard
isolation in a region
with $\met<10\,\gev$, which is dominated by multijet events. 
In the $\mu\mu jj$ selection,
the requirement of two reconstructed muons with 
large invariant mass
suppresses background from multijet production.
The background contribution from diboson or single top quark production, which
could potentially contribute in both single muon and dimuon analyses,
are found to be negligible as well.
Leptoquark pair production is simulated using \textsc{pythia} for
leptoquark masses ranging from 140\,GeV to 320\,GeV, 
corresponding to production cross sections between 2.4\,pb and 0.0074\,pb.
The generated events are processed through a full simulation of the D0
detector based on \textsc{geant} \cite{geant}.

%
%
Leptoquark candidate events are selected by requiring at least two jets with 
$p_T>25\,\mathrm{GeV}$ and muons with $p_T>20\,\mathrm{GeV}$. 
For the $\mu\mu jj$ signature at least two muons of loose quality are 
required, while for the $\mu\met jj$ selection, events with exactly one muon 
of tight quality are kept and a veto on events with additional loose muons 
is applied to ensure that the samples have no overlap.

For the $\mu\mu jj$ sample, the dimuon invariant 
mass $M(\mu,\mu)$ reconstructed from the two leading muons (i.e.\ muons with
highest $p_T$) is
required to be larger than 50\,GeV.
The numbers of $Z/\gamma^*$ and $W$ boson events 
are simultaneously normalized to data with a common scale factor 
in the region of the $Z$ boson resonance 
(defined as $60\,\gev < M({\mu,\mu}) < 120\,\gev$) after all the preceding 
cuts. At this stage, 913 data events and $930\pm 151$ expected 
background events remain, with a signal efficiency of 39.7\% for an 
assumed leptoquark mass $M_{LQ}=280\,\gev$.

\begin{figure*}
  \begin{minipage}[t]{0.5\textwidth}
    \centering
    \includegraphics[width=3.5in]{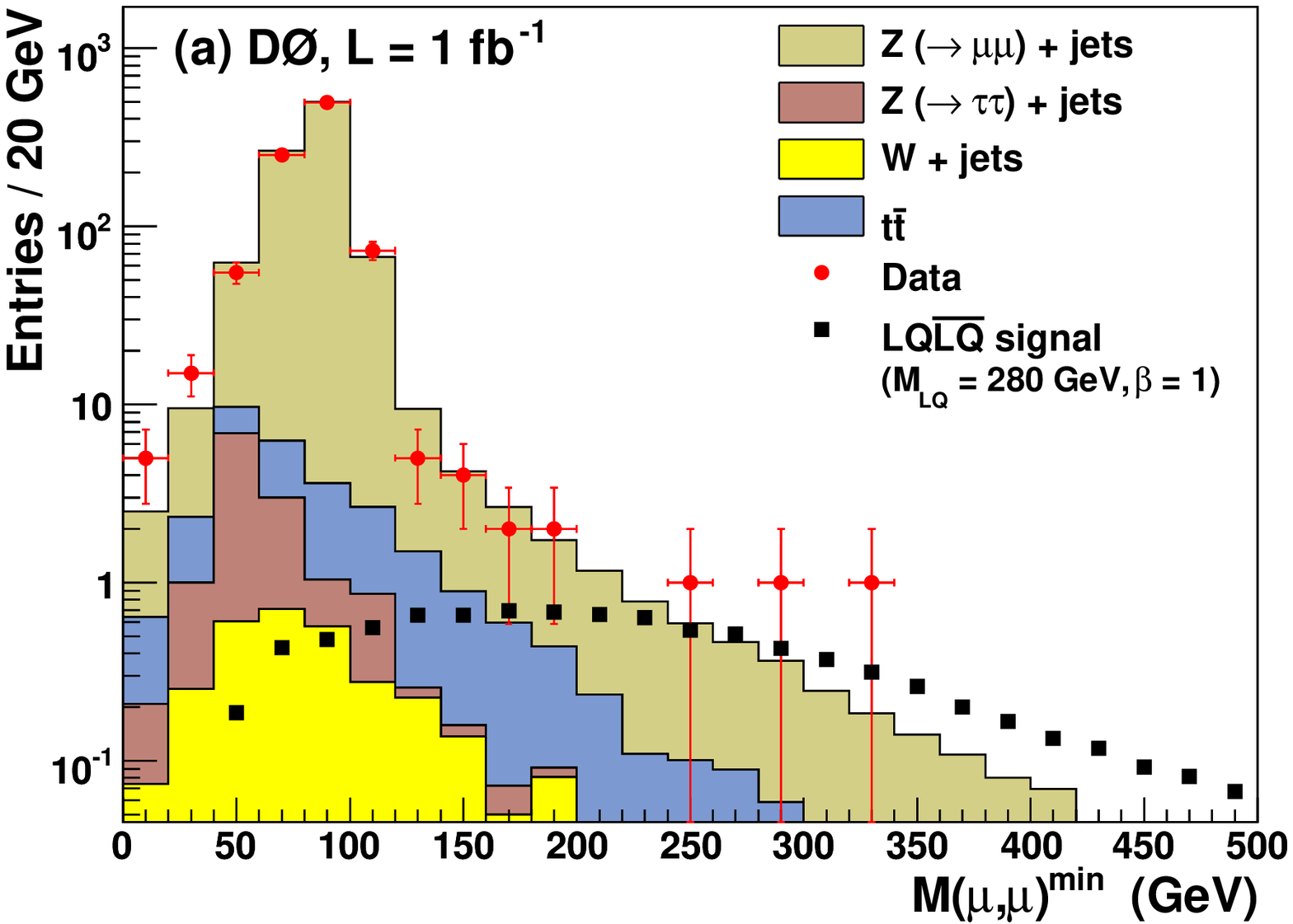}
  \end{minipage}%
  \begin{minipage}[t]{0.5\textwidth}
    \centering
    \includegraphics[width=3.5in]{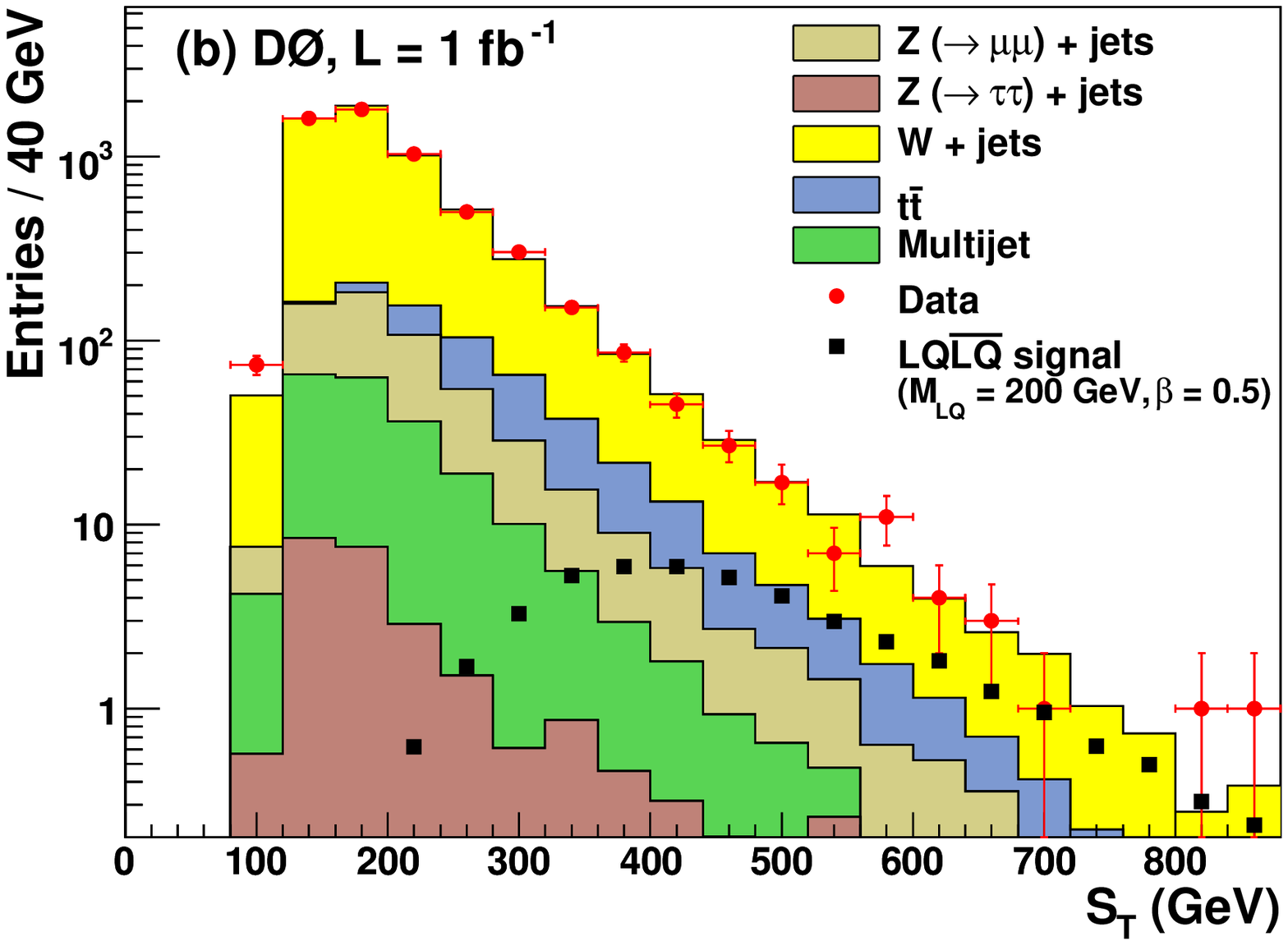}
  \end{minipage}
\caption{\label{presel_plots}  
  [color online] (a) $\mu\mu jj$ selection, minimum of the initial and 
  corrected dimuon invariant mass. 
  (b) $\mu\met jj$ selection, sum of $\met$ and the transverse momenta of the 
  final state muon and the two jets.
  Both plots show data (red dots), signal (black squares), and 
  background (colored histograms) events. Branching fractions $\beta=1$ 
  and $\beta=0.5$ and leptoquark masses of 280\,GeV and 200\,GeV 
  are assumed for the signal events in (a) and (b), 
  respectively.
}
\end{figure*}

To account for the muon $p_T$ resolution which can result in an 
overestimation of $M(\mu,\mu)$ and a mismeasurement of $\met$, and in order to 
enhance the separation between signal and background, the reconstructed 
dimuon invariant mass is corrected by 
substracting the projection of $\met$ onto the leading muon direction from 
the leading muon $p_T$. 
In this way $\met$ is minimized, which is consistent with the 
expectation that there is no genuine $\met$ in both signal 
and $Z/\gamma^*$ background. 
The minimum of the initial and corrected dimuon 
invariant mass $M(\mu,\mu)^\mathrm{min}$ is utilized as a selection 
variable to improve the rejection of $Z$ bosons (Fig.~\ref{presel_plots}).
To achieve signal enhancement and background reduction, the sum 
of the transverse momenta of the two muons and the two leading jets, 
$S_{T}=p_T(\mu_1)+p_T(\mu_2)+p_T(j_1)+p_T(j_2)$,
is also considered since the 
decay products of the SM backgrounds are likely to be less 
energetic than those of the leptoquark pair.
All combinations of the two muons with the two leading jets are 
taken to 
calculate four muon-jet invariant masses $M(\mu,\mathrm{jet})$ which are 
related to the leptoquark mass for the signal processes. Together with 
$M(\mu,\mu)^\mathrm{min}$ and $S_T$, these four muon-jet invariant masses 
$M(\mu,\mathrm{jet})$ define the set of discriminating variables that will 
be used for limit determinations.

In the case of the $\mu\met jj$ selection, the rate of multijet events is
reduced by requiring the transverse mass 
$M_{T}(\mu,\met)=\sqrt{2p_T(\mu)\met\left[1-\cos\Delta\phi(\mu,\met)\right]}$ 
reconstructed 
from the muon and the missing transverse energy to be 
larger than 50\,GeV and also by imposing $\met > 30\,\gev$. In order to remove 
events with mismeasured muon $p_T$, which could lead
to large $\met$, 
the azimuthal angle between the missing transverse energy and the muon is 
constrained to be smaller than 3.0\,radians. After applying the preceding 
cuts, the contribution from $W$ and 
$Z/\gamma^*$ boson production is simultaneously normalized to data 
with a common scale factor in the region 
$50\,\gev < M_{T}(\mu,\met) < 110\,\gev$, which is dominated by $W$ boson
production. 
At this stage, 5693 data events and $5748\pm 395$ expected 
background events remain. When assuming $M_{LQ}=200\,\gev$, the corresponding 
signal efficiencies for $\mu q \nu q$ and $\mu q\mu q$ events are 22.6\% 
and 10.2\%, respectively.

Similar to the dimuon sample, six discriminating 
variables are chosen for the single muon selection.
$M_{T}(\mu,\met)$ provides a good separation between signal and $W$ boson
events.
The large mass of the hypothetical
leptoquark motivates the choice of the following kinematic variables: 
the sum of $\met$ and the transverse momenta of the final state muon and 
the two jets, 
$S_{T}=p_T(\mu)+p_T(j_1)+p_T(j_2)+\met$
(Fig.~\ref{presel_plots}), the two 
transverse masses $M_{T}(\met,\mathrm{jet})$ constructed from the missing 
transverse energy and each of the two highest $p_T$ jets, and the two 
invariant masses $M(\mu,\mathrm{jet})$ derived from the muon 
and the two leading jets.

%
%
For each of the two selections, the six discriminating kinematic variables are 
combined into a multivariate classifier. 
A good separation between the leptoquark signal and the SM 
background is obtained with the 
{\it k\/}-Nearest-Neighbors algorithm ({\it kNN}\/), as implemented in the 
{\sc tmva}~\cite{tmva} library. 
The classification relies on the comparison of a test event
to reference events taken from training data sets.
The implemented algorithm can be interpreted as a generalization of 
the maximum likelihood classifier to {\it n} 
dimensions, where {\it n} is the number of variables used for the 
discrimination of signal against background. During the training phase, the 
classification of an event as being either signal or background is 
achieved by estimating the local signal-like probability density. This is 
defined as the ratio of the signal events over the background plus signal 
events in the vicinity of the tested event, such that the denominator, i.e. 
the number of neighbor events, is equal to the input parameter $k$. For
our analysis, the optimal value for $k$ is found to be 50, which results
in nearly maximal signal over background ratios
for reasonable computing times.
The output of the discriminant takes values between 0 and 1, 0 referring to 
the most background-like events and 1 to the most signal-like.

\begin{figure*}
  \begin{minipage}[t]{0.48\textwidth}
    \centering
    \includegraphics[width=3.5in]{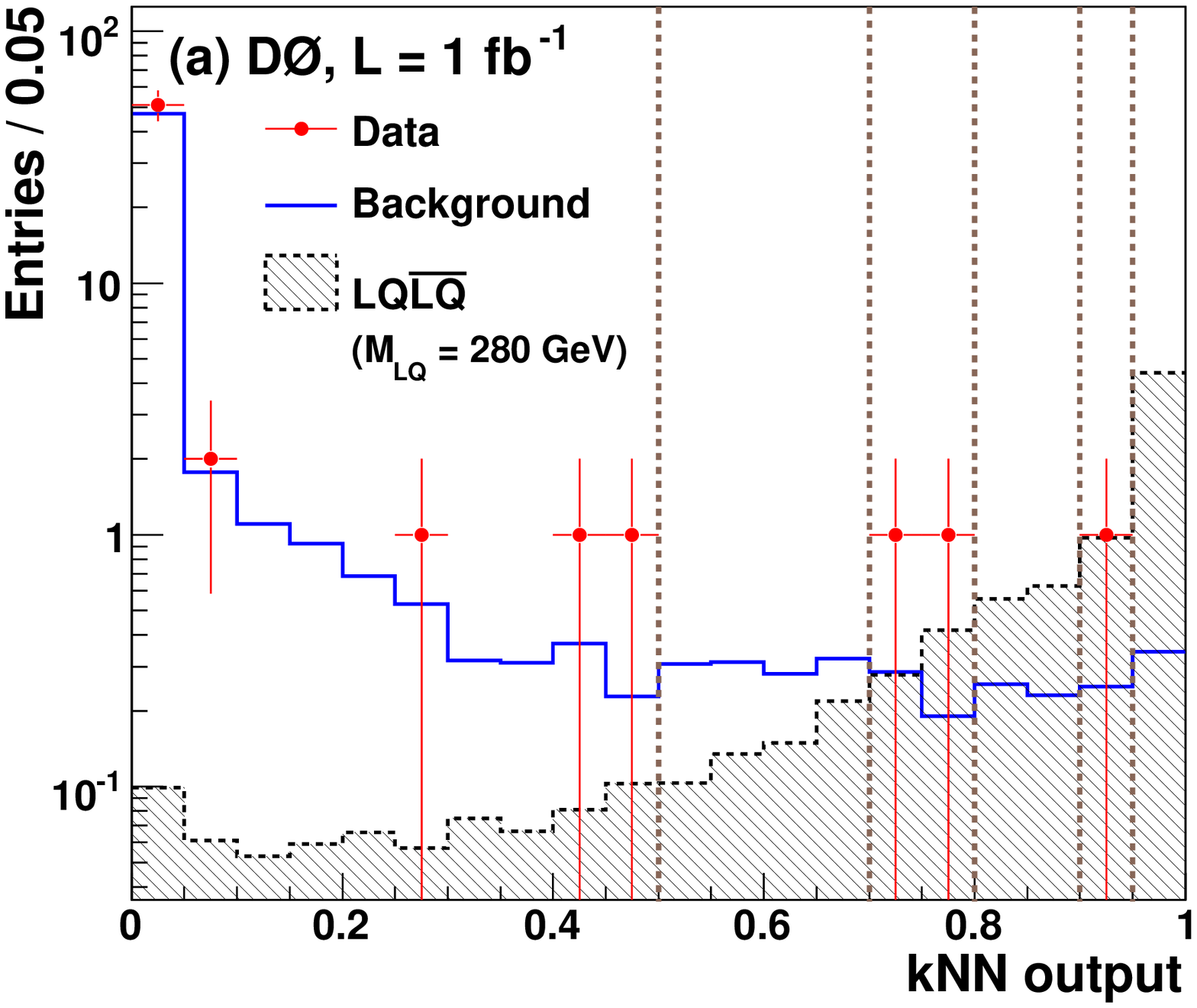}
  \end{minipage}%
  \begin{minipage}[t]{0.48\textwidth}
    \centering
    \includegraphics[width=3.5in]{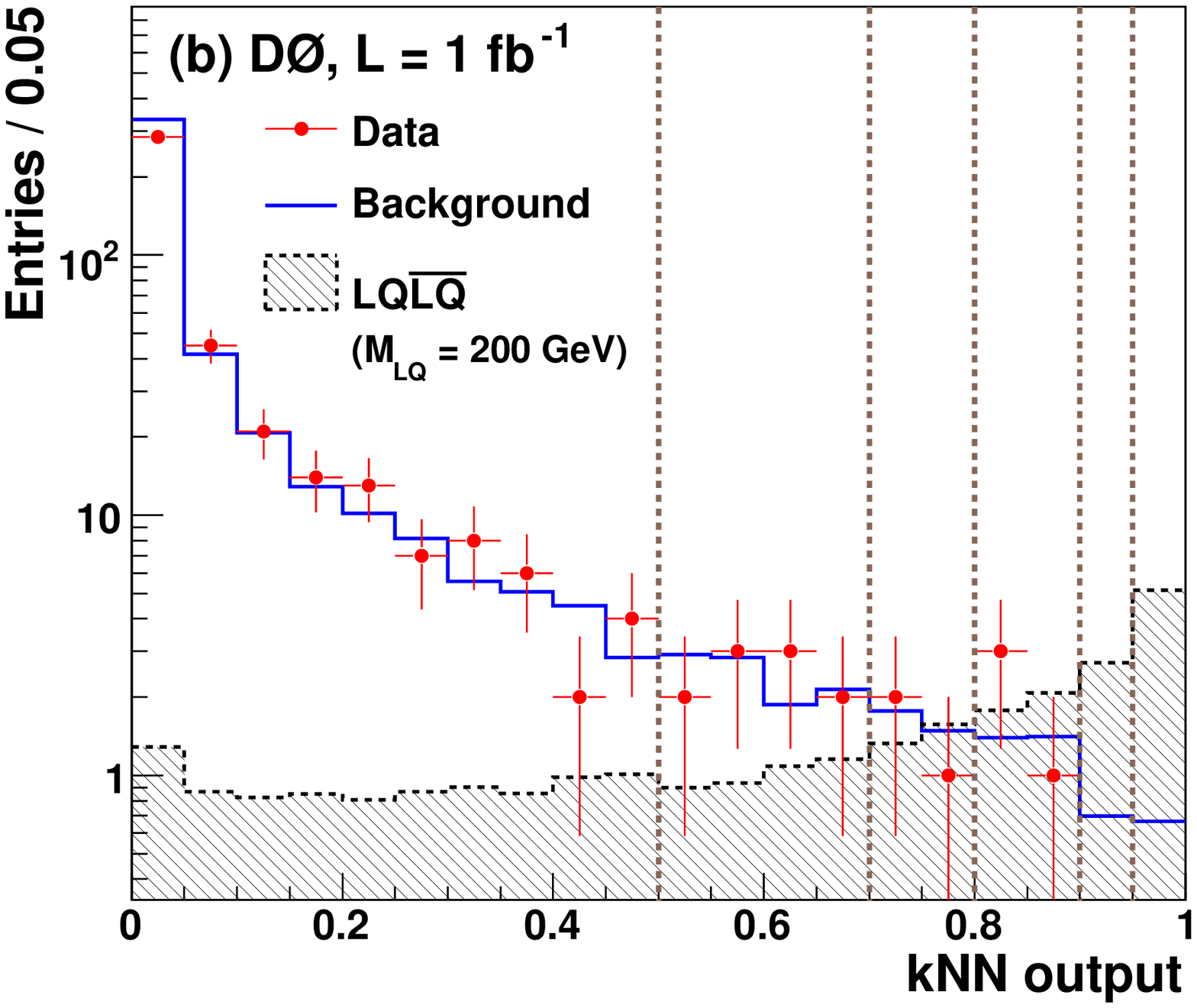}
  \end{minipage}
\caption{\label{kNN_output} 
  [color online] {\it kNN} output variable in the (a) $\mu\mu jj$
  and (b) $\mu\met jj$ selections.
  Both plots show data (red dots), signal (filled histograms), and 
  background (full lines) events. Branching fractions $\beta=1$ 
  and $\beta=0.5$ and leptoquark masses of 280\,GeV and 200\,GeV 
  are assumed for the signal events in (a) and (b), 
  respectively. The vertical lines correspond to the borders of the bins
  used in the calculation of the cross section limits.
}
\end{figure*}

The training phase is performed for each assumed leptoquark mass in the
range between 140 and 320\,GeV separately (in steps of 20\,GeV), 
and is based on simulated signal and background samples for which additional 
selections are imposed to remove regions with negligible signal contribution.
For both channels, 
the corresponding $S_{T}$ variable is required to be greater than 200\,GeV, 
while $M(\mu,\mu)$ and $M_{T}(\mu,\met)$ are required to exceed 
100\,GeV and 110\,GeV for the dimuon and the single muon analyses, 
respectively.
In the case of the $\mu\met jj$ selection, the shape of the signal event 
distributions does not vary as a function of $\beta$, as 
the topologies arising from the $LQ\overline{LQ}\rightarrow\mu q \mu q$ 
(with one muon not being reconstructed) and $\mu q\nu q$ decay modes are 
similar. 
Only one value of $\beta$ is therefore needed to complete the
training for this channel. 
In the case of the dimuon analysis, only the $\mu q\mu q$ signal events  
contribute and are considered. The performance of the training phase for the 
different assumed leptoquark masses is thus also independent of $\beta$.

In order to avoid overtraining on statistical fluctuations, 
the training phase is 
performed on half of each signal and background samples. When using the 
signal and background samples to train the {\it kNN} classifier, alternatively 
one event is kept for the training and the next one is included in a test 
sample. Good agreement in the {\it kNN} distributions between the training 
and the test samples is 
observed, which demonstrates the absence of overtraining.

%
%
Instead of cutting on the {\it kNN} variable, we choose a strategy to 
optimize the sensitivity that uses the shape of the full {\it kNN} 
distribution.
The {\it kNN} distributions are 
divided into six bins of variable size which decrease with increasing 
signal efficiency: 0--0.5, 0.5--0.7, 0.7--0.8, 0.8--0.9, 
0.9--0.95, and 0.95--1.
The same binning is used for each assumed leptoquark mass and $\beta$ 
value. The distributions of the {\it kNN} output variable are shown for both 
selections in Fig.~\ref{kNN_output}. The content of each bin is given in 
Tables~\ref{c2_kNN_bins} and 
\ref{c1_kNN_bins} for the $\mu\mu jj$ and $\mu\met jj$ 
selections, respectively.

\begin{table*}
\caption{\label{c2_kNN_bins} 
  Content of each bin of the {\it kNN} variable after preselection for the 
  $\mu\mu jj$ analysis.
  The signal efficiencies for 
  leptoquark decay into the $\mu q\mu q$ state and the expected number of 
  signal events are given, 
  as well as the numbers of events in the data and predicted background.
  The multijet background is negligible.
  The assumed leptoquark mass is 280\,GeV and $\beta$ is taken equal to 1. 
  The first uncertainties are statistical and the second systematic.
  The systematic uncertainty on the integrated luminosity is not included.
}
\begin{ruledtabular}
\begin{tabular}{l|ccc}

  \multicolumn{1}{c|}{ Samples} & 
  \multicolumn{1}{c}{ $0<kNN<0.5$} &
  \multicolumn{1}{c}{ $0.5<kNN<0.7$} &
  { $0.7<kNN<0.8$} 
  \\ \hline

$W(\rightarrow \ell\nu)+\mathrm{jets}$
 & 0.57 $\pm$ 0.12 $\pm$ 0.07
 & 0.0045 $\pm$ 0.0045 $\pm$ 0.0008
 & $-$
\\

$Z/\gamma^{*}(\rightarrow \ell^+\ell^-)+\mathrm{jets}$
 & 48 $\pm$ 1 $\pm$ 6
 & 1.00  $\pm$ 0.09 $\pm$ 0.18
 & 0.42 $\pm$ 0.04 $\pm$ 0.03
\\

$t\overline{t}$
 & 4.4 $\pm$ 0.1 $\pm$ 0.8
 & 0.22 $\pm$ 0.03 $\pm$ 0.08
 & 0.058 $\pm$ 0.018 $\pm$ 0.02
\\ \hline

Total background
 & 53 $\pm$ 1 $\pm$ 6
 & 1.2 $\pm$ 0.1 $\pm$ 0.2
 & 0.48 $\pm$ 0.05 $\pm$ 0.04
\\ \hline

{Data }
& 56
& 0
& 2
\\ \hline

{$\epsilon_\mathrm{signal}^{\,\mu q\mu q}$ (\%) }
& 2.94
& 2.47
& 2.83
\\ \hline

{$N_\mathrm{signal}$ }
 & 0.69 $\pm$ 0.02 $\pm$ 0.05
 & 0.59 $\pm$ 0.02 $\pm$ 0.02
 & 0.67 $\pm$ 0.02 $\pm$ 0.03
\\

\hline
\hline

  \multicolumn{1}{c|}{ Samples} & 
  \multicolumn{1}{c}{ $0.8<kNN<0.9$} &
  \multicolumn{1}{c}{ $0.9<kNN<0.95$} &
  \multicolumn{1}{c}{ $0.95<kNN<1$ } 
    \\ \hline

$W(\rightarrow l\nu)+\mathrm{jets}$
 & 0.0024 $\pm$ 0.0024 $\pm$ 0.0005
 & $-$
 & $-$
\\

$Z/\gamma^{*}(\rightarrow \ell^+\ell^-)+\mathrm{jets}$
 & 0.43 $\pm$ 0.03 $\pm$ 0.15
 & 0.22 $\pm$ 0.03 $\pm$ 0.02
 & 0.31 $\pm$ 0.03 $\pm$ 0.04
\\

$t\overline{t}$
 & 0.058 $\pm$ 0.016 $\pm$ 0.014
 & 0.031 $\pm$ 0.010 $\pm$ 0.009
 & 0.035 $\pm$ 0.011 $\pm$ 0.007
\\ \hline

Total background
 & 0.49 $\pm$ 0.04 $\pm$ 0.16
 & 0.25 $\pm$ 0.03 $\pm$ 0.03
 & 0.34 $\pm$ 0.03 $\pm$ 0.05
\\ \hline

{Data }  
& 0
& 1
& 0
\\ \hline

{$\epsilon_\mathrm{signal}^{\,\mu q\mu q}$ (\%) } 
& 4.81
& 3.97
& 18.0
\\ \hline

{$N_\mathrm{signal}$ } 
 & 1.13 $\pm$ 0.02 $\pm$ 0.04
 & 0.94 $\pm$ 0.02 $\pm$ 0.03
 & 4.23 $\pm$ 0.05 $\pm$ 0.22
\\

\end{tabular}
\end{ruledtabular}
\end{table*}

\begin{table*}
\caption{\label{c1_kNN_bins} 
  Content of each bin of the {\it kNN} variable after preselection for the 
  $\mu\met jj$ analysis. 
  The signal efficiencies for 
  leptoquark decay into the $\mu q\nu q$ and $\mu q\mu q$ 
  states and the expected number of signal events are given, 
  as well as the numbers of events in the data and 
  predicted background.
  The assumed leptoquark mass is 200\,GeV and $\beta$ is taken equal to 0.5. 
  The first uncertainties are statistical and the second are systematic.
  The systematic uncertainty on the integrated luminosity is not included.
}
\begin{ruledtabular}
\begin{tabular}{l|ccc} 

  \multicolumn{1}{c|}{ Samples} & 
  \multicolumn{1}{c}{ $0<kNN<0.5$} &
  \multicolumn{1}{c}{ $0.5<kNN<0.7$} &
  \multicolumn{1}{c}{ $0.7<kNN<0.8$} 
  \\ \hline

$W(\rightarrow l\nu)+\mathrm{jets}$
 & 348 $\pm$ 5 $\pm$ 44
 & 6.1 $\pm$ 0.5 $\pm$ 1.3
 & 1.8 $\pm$ 0.3 $\pm$ 0.4
\\

$Z/\gamma^{*}(\rightarrow \ell^+\ell^-)+\mathrm{jets}$
 & 53 $\pm$ 1 $\pm$ 6 
 & 1.13 $\pm$ 0.22 $\pm$ 0.15 
 & 0.40 $\pm$ 0.07 $\pm$ 0.08
\\

$t\overline{t}$
 & 34.9 $\pm$ 0.4 $\pm$ 6.5
 & 2.5 $\pm$ 0.1 $\pm$ 0.5
 & 1.0 $\pm$ 0.1 $\pm$ 0.2
\\

Multijet
 & 8.3 $\pm$ 0.3 $\pm$ 1.7
 & 0.030 $\pm$ 0.017 $\pm$ 0.006
 & 0.020 $\pm$ 0.014 $\pm$ 0.004
\\ \hline

Total background
 & 444 $\pm$ 5 $\pm$ 53
 & 9.8 $\pm$ 0.6 $\pm$ 1.7
 & 3.3 $\pm$ 0.3 $\pm$ 0.6
\\ \hline

{Data }
& 405
& 10
& 3
\\ \hline

{$\epsilon_\mathrm{signal}^{\,\mu q\nu q}$ (\%) } 
& 5.19
& 2.35
& 1.73
\\ \hline

{$\epsilon_\mathrm{signal}^{\,\mu q\mu q}$ (\%) }
& 3.57
& 1.08
& 0.517
\\ \hline

$N_\mathrm{signal}$
 & 8.9 $\pm$ 0.1 $\pm$ 0.4
 & 3.9 $\pm$ 0.1 $\pm$ 0.1
 & 2.8 $\pm$ 0.1 $\pm$ 0.2
\\ 

\hline
\hline

  \multicolumn{1}{c|}{ Samples} & 
  \multicolumn{1}{c}{ $0.8<kNN<0.9$} &
  \multicolumn{1}{c}{ $0.9<kNN<0.95$} &
  \multicolumn{1}{c}{ $0.95<kNN<1$ } 
    \\ \hline

$W(\rightarrow l\nu)+\mathrm{jets}$
 & 1.7 $\pm$ 0.3 $\pm$ 0.5
 & 0.31 $\pm$ 0.09 $\pm$ 0.24
 & 0.39 $\pm$ 0.13 $\pm$ 0.18
\\

$Z/\gamma^{*}(\rightarrow \ell^+\ell^-)+\mathrm{jets}$
 & 0.23 $\pm$ 0.04 $\pm$ 0.01
 & 0.11 $\pm$ 0.02 $\pm$ 0.01
 & 0.064 $\pm$ 0.006 $\pm$ 0.022
\\

$t\overline{t}$
 & 0.84 $\pm$ 0.05 $\pm$ 0.19
 & 0.26 $\pm$ 0.03 $\pm$ 0.05
 & 0.21 $\pm$ 0.03 $\pm$ 0.06
\\

Multijet
 & 0.030 $\pm$ 0.017 $\pm$ 0.006
 & 0.020 $\pm$ 0.014 $\pm$ 0.004
 & 0.0099 $\pm$ 0.0099 $\pm$ 0.0020
\\ \hline

Total background
 & 2.8 $\pm$ 0.3 $\pm$ 0.5
 & 0.70 $\pm$ 0.10 $\pm$ 0.25
 & 0.67 $\pm$ 0.14 $\pm$ 0.22
\\ \hline

Data              
& 4
& 0
& 0
\\ \hline

{$\epsilon_\mathrm{signal}^{\,\mu q\nu q}$ (\%) } 
& 2.28
& 1.64
& 3.15
\\ \hline

{$\epsilon_\mathrm{signal}^{\,\mu q\mu q}$ (\%) } 
& 0.718
& 0.444
& 0.600
\\ \hline

{$N_\mathrm{signal}$ } 
 & 3.7 $\pm$ 0.1 $\pm$ 0.1
 & 2.6 $\pm$ 0.1 $\pm$ 0.1
 & 5.0 $\pm$ 0.1 $\pm$ 0.3
\\ 

\end{tabular}
\end{ruledtabular}
\end{table*}

%
%
The systematic uncertainties on the predicted number of background events
and on the signal efficiencies are studied by varying the efficiencies
and resolutions for the reconstructed objects and the modeling of both
background and signal within their uncertainty range~\cite{pcthesis}. 
For each uncertainty source
and for each assumed value for $\beta$ and $M_{LQ}$,
the {\it kNN} output variable is recalculated and the deviation observed 
in each of the {\it kNN} bins used for the classification is taken as the 
systematic uncertainty. Below we quote the uncertainties obtained when 
assuming $M_{LQ}=200\,\mathrm{GeV}$ and $\beta=0.5$ for the $\mu\met jj$ 
analysis and $M_{LQ}=280\,\mathrm{GeV}$ and $\beta=1$ for the $\mu\mu jj$ 
selection. For other assumptions on $M_{LQ}$ and $\beta$, the uncertainties 
are similar.

The dominant systematic uncertainties are listed in Table~\ref{tab:syst}.
For the background they include uncertainties on the muon $p_T$ resolution,
the jet energy scale, the $t\bar{t}$ cross section, and the modeling of jet
radiation in $W/Z+$ jets events. The latter is studied by comparing the
$p_T$ distribution of the second highest $p_T$ jet observed in data
with the predictions
of the {\sc alpgen} and {\sc pythia} event generators when selecting $W$ boson
or $Z$ boson events. While {\sc alpgen} correctly reproduces the
shape observed in data, {\sc pythia} is found to underestimate the rate at
large $p_T$. A mixture of {\sc alpgen} and {\sc pythia} events, with a 30\%
(43\%) contribution of the latter, gives an acceptable description for
$W+$ jets ($Z$+jets) events with a Kolmogorov-Smirnov probability corresponding
to a $1\sigma$ variation. 

Additional uncertainties arise from the efficiencies of the muon 
trigger and the muon identification (2\%), of the jet reconstruction (0.5\%), 
and from the normalization of the $W/Z+$ jets background (2\%). The relative
uncertainty on the integrated luminosity is equal to 6.1\%~\cite{d0lumi}.
The uncertainty on the multijet background in the $\mu\met jj$ analysis
due to the extrapolation into the signal region and due to its
normalization is estimated to be 20\%.

For the signal efficiency and acceptance, the following systematic 
uncertainties are studied
in addition to the uncertainties arising from the muon and jet measurements.
The uncertainty due to the PDF choice is determined to be
1.6\%, using the twenty-eigenvector basis of the CTEQ6.1M
PDF set~\cite{cteq}. The effects of initial and final state radiation
(ISR and FSR), which 
might lead to the generation of additional jets, are studied by varying 
the {\sc pythia} parameters controlling the QCD scales and the maximal
allowed virtualities used in the simulation of the space-like
and time-like parton showers. The corresponding uncertainty on
the signal efficiencies is determined to be~1.5\%.

\begin{table*}
\caption{\label{tab:syst}
  The dominant systematic uncertainties (in \%) on the expected number of
  background events (split between $W/Z$ and $t\bar{t}$ production)
  and on the leptoquark ($LQ$) signal efficiency and acceptance 
  for the $\mu\met jj$ 
  and $\mu\mu jj$
  selections. The uncertainty range found for the six bins of each distribution
  of the {\it kNN} discriminant is quoted assuming 
  $M_{LQ}=200\,\mathrm{GeV}$ and $\beta=0.5$ for the $\mu\met jj$ selection 
  and $M_{LQ}=280\,\mathrm{GeV}$ and $\beta=1$ for the $\mu\mu jj$ analysis.
  The relative uncertainty on the integrated luminosity is 6.1\%.
  }
\begin{ruledtabular}
\begin{tabular}{l|ccc|ccc}
& \multicolumn{3}{c|}{$\mu\met jj$ channel} 
& \multicolumn{3}{c}{$\mu\mu jj$ channel}\\ 
Uncertainty & $W/Z$   & $t\bar{t}$ & $LQ$ & $W/Z$ & $t\bar{t}$ & $LQ$\\
\hline
Muon $p_T$ resolution            & $6-27$ & $4-23$ & $0.8-3.2$
                                 & $2-19$ & $0-25$ & $0.1-4.5$\\
Jet energy scale                 & $1-20$ & $0-11$ & $0.1-3.6$
                                 & $0-11$ & $0-24$ & $0.3-4.7$\\
$W/Z+$jets model                 & $2-77$ & $-$    & $-$      
                                 & $4-29$ & $-$    & $-$      \\
$t\bar{t}$ cross section         & $-$    & $18$   & $-$      
                                 & $-$    & $18$   & $-$      \\
ISR/FSR                          & $-$    & $-$   & $1.5$     
                                 & $-$    & $-$   & $1.5$      \\
PDF                              & $-$    & $-$   & $1.6$     
                                 & $-$    & $-$   & $1.6$      \\

\end{tabular}
\end{ruledtabular}
\end{table*}

\begin{table*}
\centering
\caption{\label{tab:xsectlimit}
NLO cross section~\cite{lqnlo} for scalar leptoquark pair production in 
$p\bar{p}$ collisions at $\sqrt{s}=1.96\tev$ using the CTEQ6.1M PDF 
set~\cite{cteq}, observed and expected $95\%$~C.L. cross section limits 
obtained for the $\mu\met jj$ selection assuming $\beta=0.5$ and the 
$\mu\mu jj$ selection assuming $\beta=1$, and observed limits  
for the combination assuming $\beta=0.5$ or $\beta=1$.
}
\begin{ruledtabular}
\begin{tabular}{c|l|ll|ll|ll}
\multicolumn{1}{l|}{} & \multicolumn{1}{l|}{} 
& \multicolumn{2}{c|}{$\mu\met jj$ channel} 
& \multicolumn{2}{c|}{$\mu\mu jj$ channel} 
& \multicolumn{2}{c}{Combination} 
\\

\multicolumn{1}{l|}{$LQ$ mass} & \multicolumn{1}{l|}{$\sigma_{\mathrm{NLO}}$}
& \multicolumn{2}{c|}{$\beta=0.5$} 
& \multicolumn{2}{c|}{$\beta=1$} 
& \multicolumn{1}{c}{$\beta=0.5$} & \multicolumn{1}{l}{$\beta=1$}
\\ 

\multicolumn{1}{c|}{(GeV)}
& \multicolumn{1}{c|}{(pb)} 
&\multicolumn{1}{c}{$\sigma_{\mathrm{obs}}$ (pb)}
&\multicolumn{1}{c|}{$\sigma_{\mathrm{exp}}$ (pb)} 
&\multicolumn{1}{c}{$\sigma_{\mathrm{obs}}$ (pb)}
&\multicolumn{1}{c|}{$\sigma_{\mathrm{exp}}$ (pb)} 
&\multicolumn{1}{c}{$\sigma_{\mathrm{obs}}$ (pb)}
&\multicolumn{1}{c}{$\sigma_{\mathrm{obs}}$ (pb)} 
\\ \hline

140 & 2.38    
    & 0.291    & 0.395    
    & 0.0446 & 0.0503
    & 0.127   & 0.0426   
    \\ 
160 & 1.08    
    & 0.219 & 0.204  
    & 0.0392 & 0.0357 
    & 0.129 & 0.0338    
    \\ 
180 & 0.525  
    & 0.144 & 0.144 
    & 0.0278 & 0.0257 
    & 0.0832   & 0.0223 
    \\ 
200 & 0.268  
    & 0.0770  & 0.106   
    & 0.0262 & 0.0206  
    & 0.0551  & 0.0220 
    \\ 
220 & 0.141   
    & 0.0619   & 0.0841
    & 0.0215  & 0.0177
    & 0.0484   & 0.0186  
    \\ 
240 & 0.0762  
    & 0.0540 & 0.0623 
    & 0.0202  & 0.0152
    & 0.0374  & 0.0157  
    \\ 
260 & 0.0419  
    & 0.0516 & 0.0572 
    & 0.0171  & 0.0135  
    & 0.0348   & 0.0142  
    \\ 
280 & 0.0233  
    & 0.0440 & 0.0514
    & 0.0135  & 0.0120 
    & 0.0281  & 0.00946 
    \\ 
300 & 0.0131  
    & 0.0423 & 0.0470
    & 0.0114  & 0.0107
    & 0.0255  & 0.00931 
    \\ 
320 & 0.00739 
    & 0.0398 & 0.0390 
    & 0.0100  & 0.0101  
    & 0.0227 & 0.00822  
    \\ 

\end{tabular}
\end{ruledtabular}
\end{table*}

For both selections no excess of data over the predicted background is
observed. Upper limits at the $95\%$~C.L. on the leptoquark production cross 
section $\sigma$ are
calculated for both selections separately and their combination using the
method proposed in Ref.~\cite{cls}. Since both decay modes $\mu q\nu q$
and $\mu q \mu q$ contribute in the single muon selection, limits on the
product of $\sigma$ and the branching fraction cannot be derived and thus the
cross section limits need to be evaluated for each value of $\beta$ separately.
The six bins (or twelve bins in case of the combination) 
in the {\it kNN} discriminant
are treated as 
individual channels and their likelihoods are combined with correlations of 
systematic uncertainties taken into account.
The limits are calculated using the confidence level $CL_S =
CL_{S+B}/CL_{B}$, 
where $CL_{S+B}$ and $CL_{B}$ are the confidence levels
for the signal plus background and background only hypotheses, 
respectively~\cite{cls}.

\begin{figure}
\centering
\includegraphics[width=3.5in]{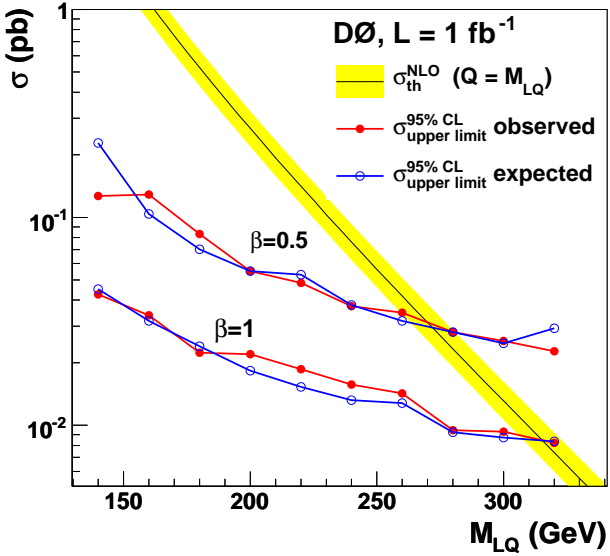}
\caption{\label{fig:xsectlimit}
[color online] Observed and expected $95\%$~C.L. upper cross section limits 
for second 
generation scalar leptoquark pair production assuming $\beta=0.5$ or
$\beta=1$. The NLO prediction is shown with an uncertainty band
reflecting the PDF and scale uncertainty.
}
\end{figure}

The observed cross section limits and the expected limits are shown
in Fig.~\ref{fig:xsectlimit} and Table~\ref{tab:xsectlimit} together with the 
theoretical prediction for scalar leptoquark pair production
calculated at next-to-leading order (NLO) in the strong coupling
constant~\cite{lqnlo} using the 
CTEQ6.1M PDF set~\cite{cteq} and the renormalization and factorization 
scale $\mu_{R,F}=M_{LQ}$.
Limits for both selections and their combination are given assuming
$\beta=0.5$ and $\beta=1$.
The uncertainty band for the cross section prediction shown in 
Fig.~\ref{fig:xsectlimit} reflects the PDF uncertainty~\cite{cteq}
and the variation of the factorization and
renormalization scale between $M_{LQ}/2$ and $2M_{LQ}$, 
added in quadrature.

\begin{figure}
\centering
\includegraphics[width=3.5in]{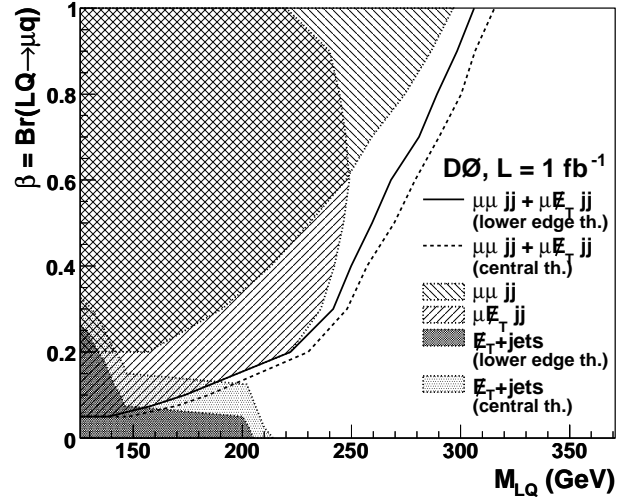}
\caption{\label{fig:masslimit}
The observed $95\%$~C.L. exclusion regions in the $M_{LQ}$ versus $\beta$ plane
obtained for the $\mu\met jj$ and
$\mu\mu jj$ selections and their combination. The exclusion regions for the
two separate selections and the solid line for the combination are obtained 
with the cross section prediction reduced by its uncertainty.
The dashed line is obtained using the nominal NLO prediction.
The exclusion at vanishing $\beta$ is based on the updated D0 search in the
acoplanar jet topology~\cite{d0lqaco2}.
}
\end{figure}

Limits on the leptoquark mass are extracted from the intersection of the 
observed upper
bound on the cross section with the NLO prediction and also the lower edge
of its uncertainty band.
Combining the $\mu\met jj$ and $\mu\mu jj$ selections and using the central 
theoretical prediction, lower bounds on the
mass of second generation leptoquarks are determined at the $95\%$~C.L. to be
$M_{LQ}>316\gev$, $M_{LQ}>270\gev$, and $M_{LQ}>185\gev$ for $\beta=1$, 
$\beta=0.5$, and $\beta=0.1$, respectively. Mass limits based on the lower edge
of the 
cross section prediction as well as the expected bounds are listed in
Table~\ref{tab:masslimit}. Figure~\ref{fig:masslimit} shows the excluded region
in the $\beta$ versus $M_{LQ}$ parameter space together with the exclusion
limits obtained for the $\mu\met jj$ and $\mu\mu jj$ selections separately.
The bound at $\beta=0$, where this analysis has no sensitivity, is given
by the D0 search in the acoplanar jet topology~\cite{d0lqaco2}.
It is interesting to notice the improvement due to the inclusion of 
the $\mu q\mu q$
decay mode in the single muon analysis. Therefore, this selection has its
maximum sensitivity around $\beta=0.6$ instead of $\beta=0.5$ and a sizable
contribution at $\beta=1$, where the branching fraction for the $\mu q\nu q$
decay mode vanishes.

\begin{table*}
\centering
\caption{\label{tab:masslimit}
Observed and expected mass limits for second generation scalar leptoquarks
assuming $\beta=1$, $\beta=0.5$, and $\beta=0.1$. The limits are
derived using the NLO prediction for the cross section or the lower edge of
its uncertainty band.
 }
\begin{ruledtabular}
\begin{tabular}{c|cc|cc}

\multicolumn{1}{c|}{}
& \multicolumn{2}{c|}{Central theory}
& \multicolumn{2}{c}{Lower edge theory}
\\

$\beta$
& \multicolumn{1}{c}{$M_{LQ}^{\mathrm{obs}}$ (GeV)} 
& \multicolumn{1}{c|}{$M_{LQ}^{\mathrm{exp}}$ (GeV)}
& \multicolumn{1}{c}{$M_{LQ}^{\mathrm{obs}}$ (GeV)} 
& \multicolumn{1}{c}{$M_{LQ}^{\mathrm{exp}}$ (GeV)}
\\ \hline

$0.1$ & 185 & 181 & 174 & 175 \\
$0.5$ & 270 & 272 & 259 & 263 \\
$1.0$ & 316 & 316 & 306 & 308 \\

\end{tabular}
\end{ruledtabular}
\end{table*}

In summary, a search for second generation scalar leptoquarks produced in
$p\bar{p}$ collisions at $\sqrt{s}=1.96\tev$ has been
performed using an integrated luminosity of $1\fbinv$. Two selections
based on the $\mu\met jj$ and $\mu\mu jj$ final states have been carried
out. The leptoquark signal was discriminated from background using a 
multivariate technique based on the {\it kNN} algorithm. Both analyses were
combined to obtain lower limits on the scalar leptoquark mass as a function of 
$\beta$ which exceed $300\gev$ at large $\beta$. 
These results improve on previous leptoquark searches at
the Tevatron~\cite{d0lq2,cdflq2}. They exceed the corresponding 
previous bounds by $55\gev$ at both $\beta=1$ and $\beta=0.5$, 
and give the most constraining direct limits on second generation
leptoquarks to date.

%
We thank the staffs at Fermilab and collaborating institutions, 
and acknowledge support from the 
DOE and NSF (USA);
CEA and CNRS/IN2P3 (France);
FASI, Rosatom and RFBR (Russia);
CNPq, FAPERJ, FAPESP and FUNDUNESP (Brazil);
DAE and DST (India);
Colciencias (Colombia);
CONACyT (Mexico);
KRF and KOSEF (Korea);
CONICET and UBACyT (Argentina);
FOM (The Netherlands);
STFC (United Kingdom);
MSMT and GACR (Czech Republic);
CRC Program, CFI, NSERC and WestGrid Project (Canada);
BMBF and DFG (Germany);
SFI (Ireland);
The Swedish Research Council (Sweden);
CAS and CNSF (China);
and the
Alexander von Humboldt Foundation (Germany).
%

%
%

\end{document}